# Multi-objective optimization: basic approaches and moving beyond them through flexible skyline queries

Giovanni Lupi

Politecnico di Milano
Milan, Italy
giovanniedoardo.lupi@mail.polimi.it

**Abstract**

The area of scientific research that deals with the simultaneous optimization of several (possibly conflicting) criteria is named multi-objective optimization. The ability to efficiently filter and extract interesting data out of large datasets is one of the key tasks in modern database systems. This paper provides a general overview of the most common approaches employed to handle the problem in the field of databases, and describes a novel framework named flexible skylines. After analyzing the main differences between single and multi-optimization problems, I will discuss the three main basic approaches used to handle multi-optimization problems: lexicographic approach, top-k queries and skylines. Each methodology will be discussed, analyzing the pros, the range of applicability and the main issues, which motivate the need to introduce the flexible skylines innovative framework. A review of this approach will show its superiority with respect to the basic approaches, as well as the capability to overcome the majority of their drawbacks.

## 1 Introduction

Multi-objective optimization is a critical area of decision making, that consists in the simultaneous optimization of different (and often conflicting) criteria. This practice has a very long history [1] and represents a typical research topic, with an extremely wide range of applicability. A vast amount of real-world optimization problems can indeed be formalized through the use of a multitude of conflicting goals. Some examples of its use include economics and finance, engineering, social sciences, and many more [2].

This paper will focus on the field of databases. Due to the ever-present connection to the Internet that characterizes the modern era, users are presented daily with the task of extracting data of interest from larger and larger datasets [3]. Determining the most interesting objects out of a given dataset can therefore be considered one of the key tasks in modern database systems [4].

A classic example is the decision process involved in buying a new car. A user might be interested in selecting the cheapest car with the best performance. Since better performance most often demands a higher price, a trade-off between these two conflicting attributes is needed.

Three main approaches have been traditionally employed to deal with the problem of simultaneously optimizing multiple conflicting criteria [5]:

1. The lexicographic approach, which enforces a priority order among the objectives

2. Top-k queries, which transform the original multi-objective into a single-objective problem
3. Skyline queries, which identify a set of non-dominated solutions

Each of these methods has notable drawbacks [3][4][5][6][7]. Motivated by such issues, I will present an overview of the notion of flexible-skyline queries [4]. This novel framework is capable of overcoming a large number of issues present in the previous basic approaches.

This paper is structured as follows.

Section 2 introduces the concept of multi-objective optimization, drawing a clear distinction with single-objective problems.

Section 3 dives into the details of each of the three traditional approaches, featuring an analysis of their pros, their range of applicability and their main issues.

Section 4 focuses on flexible skylines and the basic principles behind its functioning.

Section 5 carries out some comparisons between flexible skylines and traditional approaches.

Section 6 mentions some further developments available in the literature.

Lastly, section 7 concludes the paper.

## 2 Multi-objective optimization

Decision making can be defined as the process of selecting the best course of action from all the available alternatives [8].

In the case of single-objective optimization problems (SOO), the goal is to find a solution that optimizes a single objective function. SOO has been employed extensively throughout all fields of engineering and can be a very valuable tool to gain more insight in the nature of a problem.

However, in real world scenarios, we often deal with the problem of optimizing multiple conflicting objectives [2]. A multi-objective optimization problem (MOO) is a problem that involves the use of multiple objective functions [8]. In the area of databases, MOO has a natural interpretation: the objectives correspond to attributes and the alternatives to objects in the database [7].

Given the previous definition of SOO, one might incorrectly assume that the task of MOO is to find an optimal solution for each objective function. MOO involves a procedure quite different from that idea [9].

A simple example should suffice to clarify the problem. Let's consider the process of choosing, between a series of alternatives, the best car available in terms of price and performance. Intuitively, cars with better performance will have a higher cost, so that the two objectives are clearly in conflict with each other. Figure 1 provides an overview of the problem. Solution A provides the car with the lowest cost, but also with the worst performance. Solution B is better suited for buyers who prioritize performance at the expense of a higher cost. It's easy to see that, since the two objectives are in contrast, no single optimal solution exists. In fact, between the extreme solutions A and B, many other solutions exist, that feature a potentially optimal trade-off between the two attributes. When two solutions are compared, one achieves a better score in a specific attribute (say price) only at the cost of sacrificing another one (performance.)

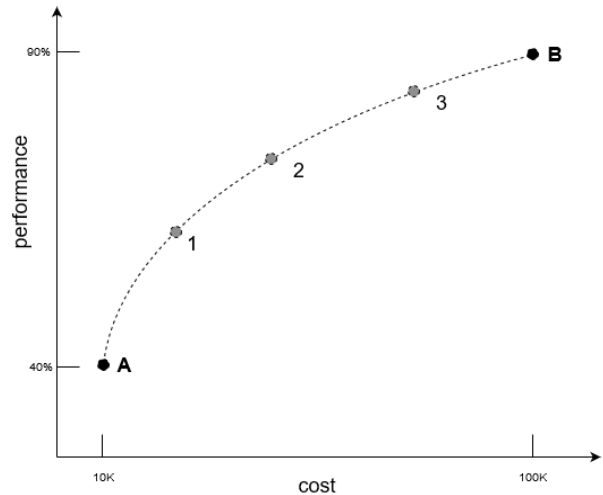

*Figure 1 – Example of Multi Objective Optimization problem*

It is clear from the example that MOO returns a set of potentially optimal solutions. Each solution is said to be Pareto optimal, as each attribute can only be improved by sacrificing another one. The set of Pareto optimal solutions is called Pareto frontier [5].

The Pareto approach will be formalized in Section 3.3 through the concept of dominance in Skyline queries.

## 3 Basic multi-objective optimization approaches

As shown in the previous section, solving MOO is not as straightforward as solving SOO. Several approaches have been designed to deal with MOO. The most popular ones in the field of databases are the lexicographic approach, top-k queries and skyline queries [5]. These can be divided in a priori and a posteriori methods [8].

A priori methods require the user to express preference information before the problem can be solved. The lexicographic approach and top-k queries belong to this category.

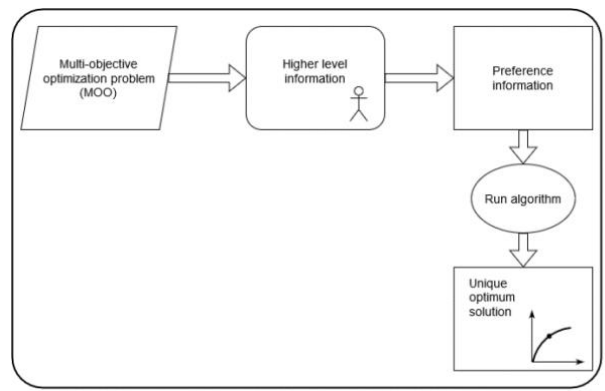

*Figure 2 - A priori method*

A posteriori methods return multiple trade-off optimal solutions and require the user to select the most satisfactory ones on the basis of non-technical and often experience-based criteria. Skyline queries are a prime example of this methodology.

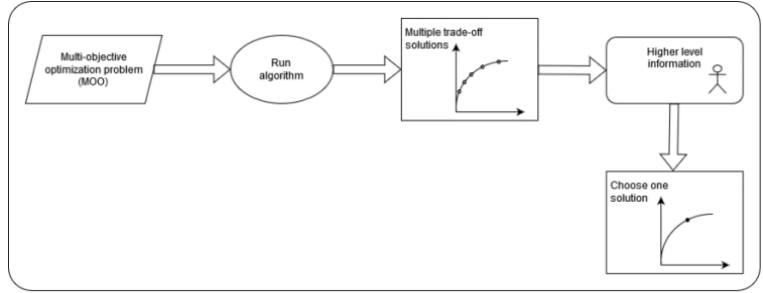

*Figure 3 - A posteriori method*

This section individually treats each approach, describing the main advantages and the field of applicability, concluding with a detailed analysis of the drawbacks.

### 3.1 Lexicographic approach

This method requires to rank each attribute in order of importance, establishing a strict priority order. The solution to a MOO is obtained by individually maximizing each criterion, according to its priority.

More formally, it is possible to order by importance each objective function that composes the MOO [8]. Let's assume that the functions are ordered from $f_1$ (highest priority objective) to $f_k$ (lowest priority objective). As a convention, I will consider lower values to be better than higher ones (the opposite convention would work the same way.) The procedure consists in solving a sequence of linear SOO of the form [10]:

$$\begin{aligned} min \quad & f_i(x) \\ s.t. \quad & f_j(x) \leq f_j(x_j^*) \quad j \in \{1, \dots, i-1\} \\ & i \in \{1, \dots, k\} \end{aligned}$$

*(1)*

In this formula, *i* represents the position of the function in the preference ordering, and $f_j(x_j^*)$ represents the solution of the $j^{th}$ optimization problem.

So, if the previous formula returns a unique solution for $f_1$, that solution is considered the optimal one for the entire problem. In the database context, this means that if a tuple is better than any other one in the dataset with regard to the highest priority objective, all the other attributes will not contribute to determine the solution [4]. If the formula instead doesn't return a unique value for $f_1$, the problem is solved for the second objective in order of importance ($f_2$). The sequence continues until the tie is broken and a unique solution is identified [5].

The lexicographic approach is an a priori method. In fact, the algorithm requires the user to specify the order of importance of each attribute before its execution. Whenever it is possible to identify a strict priority among the attributes, this algorithm represents a good choice. It has several advantages. First, it is simpler than the other popular methods. Also, unlike top-k queries, this method treats each individual criterion separately, without the need to mix measures related to very different attributes of the solution.

The need to specify a priori a strict preference order for the attributes is however a big drawback of this approach. In fact, in many real-world scenarios, identifying a strict preference order is not possible. The approach is also severely lacking in terms of flexibility: a small difference in the preferred attribute completely outweighs any (potentially wide) difference in the other ones. Lastly, it requires to solve many SOO to obtain a single solution.

### 3.2 Top-k queries

Top-k queries (or ranking queries) are currently the most popular approach to handle MOO [5][9]. The idea behind them is to transform the underlying MOO into a SOO by means of a weighted formula, in order to retrieve the best *k* solutions to the problem. This formula is referred to as scoring function (or ranking function.)

Going into greater detail, the user assigns to each attribute ($a_i$) a numerical weight ($w_i$), representing its priority. The set of objectives is then scalarized into a single objective by means of a weighted-sum. Assuming minimal values are preferred, the problem is reduced to minimizing and extracting the *k* best results from an objective function of the form:

$$F = a_1 w_1 + a_2 w_2 + \cdots + a_n w_n$$

(2)

where *n* represents the number of relevant attributes.

Ranking queries are an a priori method. The user is requested to input a preference vector ***w*** that will be used to construct the unique objective function in charge of solving the problem. If such a relative preference among the attributes is known a priori, this is an ideal method and there is likely very little reason to choose a different one [9]. In fact, this method is quite simple and allows to retrieve a set of optimal solutions by solving only one SOO. Using a monotone scoring function (as most of the popular top-k processing techniques do) allows to further increase the efficiency of the computation [11]. Additionally, this method offers a fine-grained control over the cardinality of the result set, since parameter *k* is customizable by the user.

On the other hand, the majority of times, selecting an accurate preference vector ***w*** is by no means a straightforward task. The reader should indeed pay attention to the fact that the results achieved by this method are extremely sensitive to the weight assigned to each attribute. A small change in the preference vector might easily result in a drastic change in the problem solution. The biggest drawback of top-k queries is the need to provide an accurate preference vector without any knowledge of the consequences and with no possibility to predict the changes caused by modifying one or more of its parameters. Review [5] shows some other interesting arguments against the use of this method. First, by using a single objective function, the algorithm misses the opportunity to find other solutions that might be more interesting to the user, representing a different trade-off between the attributes. Second, the scalarization might cause the need to add quantities that use very different units of measurements, therefore requiring to normalize the value of each attribute within a certain range. Lastly, even after applying the normalization, one should question the meaning of adding attributes that might very well be non-commensurable. Regardless of the normalization, combining attributes so different from each other might not be meaningful, ultimately defeating the purpose of retrieving data of interest.

### 3.3 Skyline queries

Skyline queries are implemented by means of a multi-objective optimization algorithm based on the concept of Pareto dominance. A tuple *t* is said to dominate a tuple *s* (indicated as $t \prec s$) if *t* is at least as good as *s* for all the attributes, and *t* is strictly better than *s* for at least one attribute [12]. Equivalently, given *n* attribute values ($a_i$), and assuming lower values are better, the following two conditions have to be satisfied:

$$\forall a_i, i = 1 \ldots n, t(a_i) \leq s(a_i)$$

(3)

$$\exists a_j, j = 1 \ldots n, t(a_j) < s(a_j)$$

(4)

A skyline query returns the set of tuples that are not dominated by any other tuple. This set is also called Pareto frontier, while each solution in the set is said to be Pareto-optimal. The Pareto frontier has the property of dominating all the other solutions that don't belong to the set [9]. Given any two Pareto-optimal solutions instead, none of them dominates each other.

Figure 4 shows the concept of Pareto dominance applied to the previous car example. To abide by the standards used in the literature, the cost and performance attributes have been normalized in the 0 to 1 range, in such a way that lower values are better. Solution A dominates solution D: in fact, A has lower values than D for both the attributes considered. Neither solution A nor B instead dominates each other. These tuples simply represent a different trade-off solution that corresponds to a different order of importance of the attributes.

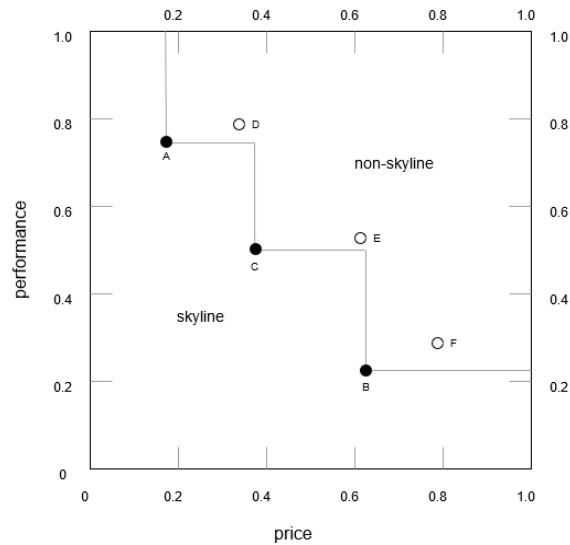

*Figure 4 - Example of skyline query*

Reference [7] also shows that the skyline can be equivalently defined as the set of tuples that minimize at least one monotone scoring function. Therefore, the skyline represents the set of potentially optimal tuples.

Skyline queries are an a posteriori method. In fact, the algorithm first returns a set of potentially optimal solutions and successively the user will select one on the basis of higher-level information. It is worth highlighting the difference with respect to the previous approaches: the information provided by the user is not used to find a new solution, as in the a priori approaches, but only to select one from a set of already obtained trade-off optimal tuples. Both approaches require the user to take a decision. In fact, as survey [5] correctly points out, participation of the user is fundamental in the process of discovering data of his interest. However, a priori decisions are often uninformed ones and can influence the results in an unpredictable way, while having a set of already given trade-off optimal solutions provides the user with better means to take an informed decision. The capability of running without specifying a preference vector and returning a rich set of potentially optimal solutions are the biggest advantages offered by this method. Additionally, unlike other popular approaches (namely the lexicographic one), it does not require multiple runs to return a result.

Unfortunately, even this approach suffers some relevant drawbacks. First, the procedure is rather complicated and with an intrinsically high computation requirement (quadratic), and becomes quite inefficient when applied to large datasets. An equally important problem is that skyline queries offer no control with respect to the cardinality of the result. In fact, they will often return an overwhelming amount of potentially optimal solutions, making impracticable for a user to select a unique one. Lastly, a skyline query always returns the same result to every user. This is a direct consequence of the inability of a posteriori methods to personalize their result. While this lack of subjectivity might be ideal in certain scenarios, it can also be an undesirable property, especially when the result is a set of high cardinality. In fact, keeping track of the user preference is what allows to filter out some undesirable tuples and reduce the size of the output. This is one of the core ideas behind flexible skylines, which will be discussed in the next section.

# 4 Flexible skylines

The shortcomings of the traditional algorithms introduced in the previous section have in recent times motivated the scientific community to design innovative approaches. Because of the multiple advantages offered by skyline queries, many of these novel approaches focus on new ways to process them [13]. Flexible skylines (F-skylines) represent an interesting development in this direction.

This new framework was originally introduced in 2017 by authors Ciaccia and Martinenghi in [6], under the name of restricted skyline queries. In 2020, the same authors extended their original paper with a series of improvements, rebranding restricted skyline queries as flexible skyline queries [4].

This section will present the main ideas underlying this methodology.

### 4.1 Rationale

One of the main drawbacks of basic skyline queries is the often-overwhelming output size they produce. Different solutions have been proposed in the literature to overcome this problem. However, the majority of these methods try to retain the most interesting points belonging to the original skyline on the basis of a general criterion, without taking into account the preference of a specific user [3].

Flexible skylines, instead, can capture the preference of a user by merging some of the core principles of ranking queries into skyline queries. As shown in section 3.2, one of the main disadvantages of top-k queries is the need to accurately specify a priori a weight vector $w$. The result is in fact extremely sensitive to the chosen weights: a small change in the vector can drastically alter the output of the algorithm. Since being able to somehow accurately mine a priori a weight vector with a high level of precision is an unreasonable expectation, it is necessary to allow some form of flexibility when considering the user preference. Flexible skylines relax the input requirements, modelling user preferences by means of a set of constraints on the weights used in a scoring function.

Thus, depending on the user preference, F-skylines are capable of focusing on a specific portion of the overall skyline. As a consequence, the size of the output will also be reduced. In particular, a tighter set of constraints results in an output of lower cardinality.

### 4.2 Preliminary concepts

Section 3.3 introduced the concepts of skyline queries and Pareto dominance. On the basis on these notions, it is convenient to define a skyline operator (SKY). Let's consider a relational schema $R$. An instance over $R$ is a set of tuples over $R$: we use the letter $r$ to refer to this instance. Given two tuples $s$, $t$ over $R$, the skyline of $r$ can be defined as:

$$SKY(r) = \{t \in r \mid \nexists s \in r.\ s \prec t\}$$

*(5)*

The SKY operator can equivalently be defined as the set of optimally potential tuples [7]. These are all the tuples that are strictly better than all the others according to at least one function in the set $MF$ of all monotone functions. In formula:

$$SKY(r) = \{t \in r \mid \exists f \in MF. \forall s \in r.\ s \neq t \rightarrow f(t) < f(s)\}$$

*(6)*

It is possible to modify the concept of dominance by considering only a limited set of monotone scoring functions. Let's consider a set $\mathcal{F}$ of monotone scoring functions. A tuple $t$ is said to $\mathcal{F}$-dominate a distinct tuple $s$ (which we denote $t \prec_{\mathcal{F}} s$) if:

$$\forall f \in \mathcal{F}.\ f(t) \leq f(s)$$

(7)

### 4.3 Flexible skyline operators

By means of the previous definition of $\mathcal{F}$-dominance, it is possible to introduce a new skyline operator, named non-dominated flexible skyline (ND). This set contains all the non-$\mathcal{F}$-dominated tuples in $r$; in formula:

$$ND(r; \mathcal{F}) = \{t \in r \mid \nexists\, s \in r.\ s \prec_{\mathcal{F}} t\}$$

(8)

Clearly, this set coincides with SKY when $\mathcal{F}$ is equal to the set $MF$ of all monotone scoring functions. In this specific case, formulas (5) and (8) become indeed identical.

It is also possible to introduce a second additional operator, named potentially optimal flexible skyline (PO). This returns the tuples that are the best according to at least one scoring function contained in $\mathcal{F}$, in formula:

$$PO(r; \mathcal{F}) = \{t \in r \mid \exists f \in \mathcal{F}.\ \forall s \in r.\ s \neq t \rightarrow f(t) < f(s)\}$$

(9)

Once again, this equation becomes identical to the definition of the SKY operator (6) when $\mathcal{F}$ coincides with $MF$.

As a consequence of the previous definitions, we can notice that in the limit case where $\mathcal{F}$ is the family of all monotone functions $MF$, ND and PO coincide:

$$PO(r; MF) = ND(r; MF) = SKY(r)$$

(10)

However, in the general case, the behavior of the two operators differs and the following relationships hold:

$$PO(r; \mathcal{F}) \subseteq ND(r; \mathcal{F}) \subseteq SKY(r)$$

(11)

A last interesting property to consider is that, quite intuitively, the stricter the set of constraints applied is, the more the cardinality of the ND and PO sets is reduced. In other terms, ND and PO are monotone operators with respect to the set of scoring functions. Therefore, for any two sets $\mathcal{F}_1$ and $\mathcal{F}_2$ such that $\mathcal{F}_1 \subseteq \mathcal{F}_2$, the following relationships hold:

$$PO(r; \mathcal{F}_1) \subseteq PO(r; \mathcal{F}_2)$$

(12)

$$ND(r; \mathcal{F}_1) \subseteq ND(r; \mathcal{F}_2)$$

(13)

Consequently, the performance of the algorithm is tightly related to the selectivity of the set of constraints: the effectiveness of an F-skyline query tends to improve as the number of constraints increases.

## 5 Comparison with the basic approaches

The lexicographic approach requires to establish a strict priority order among the attributes. Its main drawback is the lack of flexibility: a small difference in the preferred attribute completely outweighs any difference in the other ones. Some hybrid solutions exist in the literature that try to combine the principles of the lexicographic approach and skyline queries, leading to the so-called prioritized skylines (p-skylines) [14]. P-skylines share the same flexibility problems of the lexicographic approach. On the opposite, F-skylines allow a tradeoff between the attributes, leading to a much higher degree of flexibility.

One of the main issues with top-k queries is the need to accurately specify a priori a preference vector. As previously shown, F-skylines relax this requirement, modelling preferences by means of a set of constraints, allowing once more for greater flexibility. Furthermore, instead of returning the optimal solution according to a single objective function, F-skylines return a set of trade-off optimal solutions, thanks to the integration with skyline queries. On the other hand, this increased capability to discover data of interest comes at the cost of a much higher computational burden, as skyline queries are intrinsically more complex than ranking queries.

Skyline queries offer no control with respect to the cardinality of their output, which can often become of impracticable size. Moreover, they lack the capability of tailoring the result to the needs of a specific user. F-skylines allow instead to keep track of preferences, which cause in turn a reduction in the size of the output. Clearly, the cardinality of the result decreases as the set of constraints applied becomes tighter. Estimating the size of a skyline query is a rather difficult task and constitutes an open research topic [15]. However, the authors in [4] carry out an interesting empirical analysis on the ratio of points retained by the ND and PO operators with respect to SKY. All their experiments display a significant reduction in the number of points. PO in particular proves to be very effective, even when the number of constraints employed is limited. Nevertheless, unlike top-k queries, F-skylines don't allow to control precisely the number of results. Lastly, paper [4] also examines the execution times of ND and PO. The time required to compute SKY and ND appears to be comparable, and actually often in favor of the second. The computation of PO instead seems to only incur in a moderate overhead. Given the small differences in terms of execution time and the simultaneous effectiveness in reducing the output size, F-skyline operators can safely be used in place of SKY.

# 6 Related work

Flexible skylines are capable of returning all the tuples in a dataset that are the top-1 according to a specific scoring function by means of the PO operator. Therefore, a possible extension to the framework could allow to generalize the notion of potential optimality for the case k ≥ 1.

Mouratidis and Tang implement this extension in [16]. Their work is based on the concept of k-skybands, a natural generalization of skylines that returns the set of tuples dominated by fewer than k others. On this premise, the authors introduce the r-skyband algorithm. This algorithm, similarly to f-skylines, allows the users to define their preferences in a flexible way. The inaccuracies in the weights' specification are handled by expanding the input vector to a region $R$ into the preference domain. In the more general case, this region forms a convex polytope. This idea is very similar to the one behind flexible skyline, where the set of linear constraints on the weights defines a convex polytope as well. Both the approaches are thus regarded as fixed-region techniques. Given a preference region $R$ and a monotone scoring function $S$ linear to the weights in $w$, a tuple $t_1$ is said to r-dominate a tuple $t_2$ if $S(t_1) \geq S(t_2)$ for any weight vector in $R$, and there is at least weight vector in $R$ for which $S(t_1) > S(t_2)$. The r-skyband of a dataset is made by the tuples that are r-dominated by fewer than $k$ others. Since the r-skyband is a subset of the traditional k-skyband, the algorithm is successful in reducing the output size.

The authors in [3] identify three hard requirements for discovering data of interest in a multi-objective scenario:

- Personalization. The algorithm should keep track of a user's preference and return a result tailored to his needs.
- Controllable output size. An accurate control of the output size can be critical to comply with certain design choices, as the target device's display size, its performances, and so on. Moreover, according to Hick's law [17], the amount of data presented to the user strictly influences the quality and the time involved in decision making. Therefore, the algorithm must be able to control exactly the number of tuples in the result. Any operator with this property is called output-size specified (OSS).
- Flexibility in the preference specification. As extensively argued in the previous sections, given the impossibility to accurately mine a priori a weight vector, the input preference requirements must be relaxed.

Flexible skylines and r-skybands are very effective in producing personalized results on the basis of relaxed input preferences. However, these approaches lack the OSS property: both algorithms cannot determine nor predict the size of their output. This drawback is common to all fixed-region techniques, where the size of the preference region $R$ is established in advance. Since the OSS property cannot be satisfied by directly extending the previous approaches, the authors devise a new strategy. Analogously to top-k queries, the algorithm allows the user to specify a weight vector $w$, thus achieving personalization. However, to allow flexibility, $w$ is only treated as a best effort estimate of the user's preferences. In fact, vector $w$ is left free to expand equally in all directions. Hence, the preference region effectively becomes a hyper-sphere centered in $w$, as opposed to a convex polytope as in the fixed-region techniques. Lastly, to achieve the OSS property, the radius of this hyper-sphere is gradually increased until the number of tuples in the result matches exactly the required size.

# 7 Conclusion

This work has presented a general overview of the multi-objective optimization problem in the field of databases. Firstly, I have introduced the traditional approaches employed to solve the problem, highlighting their main pros and cons. Next, I presented flexible skyline queries, which represent a significantly more advanced technique. This method is capable of overcoming the majority of the disadvantages of the basic techniques. Given the numerous advantages offered by f-skylines, they can safely be used in place of regular skyline queries. To conclude, I analyzed some related articles that further improve or extend the basic principles behind flexible skylines.